 \documentstyle[11pt,aaspp]{article}
\newcommand{\BE}{\begin{equation}}
\newcommand{\BEAL}{\begin{eqnarray}}
\newcommand{\EE}{\end{equation}}
\newcommand{\EEAL}{\end{eqnarray}}
\def\_#1{_{\scriptscriptstyle #1}}
\def\^#1{^{\scriptscriptstyle #1}}
\def\ecmt{\rm{erg}\,\rm{cm}^{-2}}
\def\kmsq{\rm{km}^2}
\def\eV{\rm{eV}}
\def\kpc{\rm{kpc}}
\def\Mpc{\rm{Mpc}}
\def\Bp{B\_{\perp}}

\def\rl{r\_{L}}
\def\dc{d\_{c}}
\def\Dt{\Delta t}
\def\dt{\delta t}

\def\a{\alpha}

\def\deg{\^{o}}
\def\lII{\ell\^{II}}
\def\bII{b\^{II}}
\def\U{UHECR}
\def\tte{$10^{20}\,{\rm \eV}$}
\begin{document}

\title{Possible association of ultra-high-energy cosmic-ray events with
strong gamma-ray bursts}
\author{ Mordehai Milgrom and Vladimir Usov}
\affil{Department of Condensed-Matter
 Physics, Weizmann Institute of Science  76100 Rehovot, Israel}
\begin{abstract}
 We point out that each of the error boxes of the two highest-energy
 cosmic-ray shower events known, overlaps with that of a strong gamma-ray
 burst (GRB). The GRBs precede the cosmic rays by 5.5, and 11 months
 respectively. In one case the strongest known cosmic ray is paired with
 the strongest gamma-ray burst in the BATSE catalogue. The probability of
this to have occurred by chance seems rather small. Without building on
 post-factum statistics, we think the above is remarkable enough to
 suggest that the cosmic ray and gamma-ray burst were produced by
 the same outburst. A time delay  (and a small positional disparity)
 is expected, since the trajectory of a charged cosmic-ray
 particle is wriggled by intervening magnetic fields. We estimate that
the Galaxy's field alone may produce a delay of the order observed.
With similar energies going into gamma rays and cosmic rays, we expect
only a few tens percent of the strongest GRBs to be followed by a
 detection of an ultra-high-energy cosmic ray in existing experiments.
 We discuss some of
 the implications that follow if such an association is confirmed. For
 example, the upper limit on the distance of the cosmic-ray source,
 combined with a much-better-determined position to the gamma-ray burst
 source, narrows greatly the volume in which to look for an optical
 counterpart. The need to produce ultra-high-energy cosmic rays in the
 GRB source imposes additional constraints on the physical conditions in
 it. There is also useful information in the time delay regarding, e.g.,
 intergalactic magnetic fields.
\end{abstract}

\section{Introduction}
\par
The nature of the events that engender cosmic gamma-ray bursts (GRB)
has not been established yet; even their rough whereabout on the cosmic
map is still a moot issue (for a review, see \cite{hrt94}).
 While various constraints have been devised
regarding physical characteristics of the underlying process, no definite
picture has emerged. As is the rule in astronomy, important further clues
would be come by if the GRB source could be identified as a generator of
a different type of signals (e.g. of electromagnetic
 radiation in the optical region).
A similar situation exists regarding the (very few)
ultra-high-energy cosmic-ray (\U) events of energy  higher than \tte.
The situation concerning optical identification of the sources of
\U s, some constraints on their origin, and physical conditions
in their sources have been recently summarized by Sigl, Schramm \&
Bhattacharjee (1994), and by Halzen et al. (1995).
\par
We have set to check whether a connection can be made between these two
types of phenomena. Expecting a time delay, we looked for
 GRBs that precede \U\ events by at most a year or two, and that come
 from the very near vicinity.
If the association we conjecture here is confirmed
 the constraints that apply to each phenomenon
 separately can be cast on the other.

\section{Cosmic-ray-GRB association}

\par
In looking for associations with GRBs we have concentrated on
only the most energetic \U s, for the following reasons.
The higher the energy of a charged cosmic ray
(CR), the less its trajectory is
affected by intervening magnetic fields,
 and the more practicable it is to
associate it with a GRB: First, trajectory wriggling  causes the CR
to appear from a false position. Sigl, Schramm \& Bhattacharjee (1994)
estimate the disparity in angle to be at most a few degrees for the most
energetic \U s
($E\sim 3\times 10^{20}\,\eV$)
(see also \S 3),
but the expected deflection angle scales in inverse proportion
to the energy. Second,
since we are interested in correlations with brief, temporal events
we must also reckon with the fact that trajectory wriggling
produces a time delay in the arrival time of the CR.
We discuss this in \S 3, but note here only that for the
 very-highest-energy CRs
 the expected time delay is already of the order of a few months,
at the minimum,
and the delay scales as the inverse square of the energy (and is very
uncertain). For lower-energy CRs there may be too long a time stretch
in which to look for candidate GRB's with any degree of confidence,
without refined statistical tools. Third,
because the energy of \U s degrades quickly with propagation through
the intergalactic medium (e.g. \cite{psb76,hs85,ssb94}) the
 highest-energy ones are probable to come from nearer by. A
GRB that comes from the same source is then expected to be stronger
(of higher flux), thus rarer among the GRB's, and, in turn,
 more confidentially
associated with the \U. In general, such more energetic events are
also better pin-pointed on the sky in both the CR and gamma-ray regime
because of better statistics. (For strong GRBs there is added bonus
in that they have a better chance of being detected
by the higher-energy instruments such as EGRET, with an even
better angular accuracy.)
\par
There are very few \U s reported in the literature with energies
above \tte .
 In the pre-1991 era--for which we have not had available
systematic GRB data to look in--there is the event detected by the
Yrkutsk experiment, in May 1989
with energy of $(1.2\pm0.4)\times$ \tte (\cite{efim91}),
and 4 events from the Haverah-Park experiment described in
(\cite{bce85}) -- which reference has not been available to us
 (their existence is reported in \cite{lrw91}).
We are then left with only two \U\ events observed from 1991 onward:
The most energetic event ever detected  was seen by the Fly's Eye
experiment on October 15 1991, with $E=(3.0\^{+0.36}\_{-0.54})\times
 10^{20}$ \eV\ (\cite{bcd93,bcd94}). The position and error box
($\a=85.2\deg\pm1\deg,~\delta=48\deg\pm10\deg$) are
shown in Fig. 2 of \cite{ssb94} in galactic coordinates.
 The second most energetic \U\ was detected by the AGASA experiment
 on December 3 1993, with $E=(1.7-2.6)\times 10^{20}$ \eV,
 and position: $\lII=131\deg,~\bII=-41
\deg$, with an error circle (1-$\sigma$) of $1\deg$ (\cite{yhh95}).

\par
We searched
the BATSE catalogues (\cite{batI,batII})
 for GRBs observed before each of the
 two \U\ events, and  within a few degrees of its reported position.
\par
 For the Fly's Eye event (hereafter FEE)
 we found within the error box a single BATSE GRB (910503)
 that was observed about 5.5 months earlier (the BATSE record
starts only about a fortnight earlier yet). This event happens to be the
highest-total-fluence event in the BATSE catalogue,
 and in all
 respects it is among the very few brightest GRBs observed to date.
It is the 1991 May 3 event that was detected also by COMPTEL
 (\cite{wbb92a}), by EGRET (\cite{sbf92,hdm94}), and by
Ulysses, and PVO (\cite{batI,vbb92}).
 A maximum photon energy of 10 G\eV\ is reported
(\cite{hdm94}). The best single-instrument
 position comes  from EGRET (\cite{sbf92}): $\lII=171.9\deg\pm1.3\deg$,
$\bII=5.3\deg\pm1.1\deg$. That of COMPTEL is
 (\cite{wbb92a}): $\lII=171.8\deg,~\bII=6.4\deg$, with an error radius of
about $3\deg$, all errors are for 99\% confidence level.
 Time delay measurements
with PVO and/or Ulysses can give a much better angular definition.
Indeed for GRB 910503 \cite{vbb92} cite a combined COMPTEL-Ulysses
 error box of only
$1\deg\times1'$, which, we find, overlaps with the \U\ error box.
 The different error regions are depicted in Fig. 6 of
 \cite{wbb92b}. The error box of the FEE
 covers a fraction of about $10^{-3}$ of the sky, and
 the chance probability to find in that box
 at least one of the $\sim$85 BATSE events
that preceded the cosmic ray is small, to say nothing of the probability
to find a rare event such as GRB 910503.
\par
 As to the AGASA event,
 in its near
 vicinity (a few degrees)
 we found no GRB among the 260 in the first BATSE catalogue (recording
starts 21 months before the CR event). In the second BATSE catalogue
(\cite{batII}) we found three GRBs: 1. GRB 920615B (June 15 1992)
 at $\lII=129.2\deg,~\bII=-39.62\deg$.
It has no quoted fluxes, but the rather large quoted photon-counting
positional error of $4.15\deg$ indicates that it was not energetic.
2. GRB 920617C (June 17 1992--perhaps related to the previous)
 from $\lII=131.18\deg,~\bII=-41.25\deg$, is also
energetically inconspicuous (photon counting positional error of
 $5.59\deg$). 3. GRB 921230 (December 30 1992--11 months before the
 AGASA event), with
 a position $\lII=132.91\deg,~\bII=-42.87\deg$ (photon-counting
positional error of $0.4\deg$) is about $3\deg$ away from the center
of the CR error box.
While the GRB statistical position error in this case is small,
BATSE is said to have a systematic uncertainty of $4\deg$
(\cite{batI,batII}). Unlike
the other case we do not have here the benefit of a more accurate
EGRET or COMPTEL position, so the association remains
even more tentative. Energetically, GRB 921230
 is not in the same league as GRB 910503
(it has a total BATSE fluence
about seven times smaller) but is definitely among
 the top ten percent in fluence.
\par
Pitfalls of post-factum statistics abound in the present case. We
 believe, however, that the facts as described above cogently bring to
mind the possibility that the \U\ event, and the GRB are engendered
by the same cataclysmic event.
At any rate, the exact statistical significance of the association need
not be established at this point. More correlations such as those above
are expected for future pairs. In fact, some strong GRBs that have been
observed recently (e.g. \cite{hdm94})
 may be good candidate precursors for future
\U\ events.

\section{Propagation effects}

\par
\U s are almost certainly extragalactic. Proton seems to be favoured
as the primary particle that triggers the event (\cite{ssb94}), but
a neutrino cannot be excluded as an atypical primary
(\cite{ssb94,hvs95}). We assume the former in what follows presently;
 neutrinos will be discussed at the end of this section.
\par
Two effects on an \U\ proton traveling though the
galactic and extragalactic medium have been extensively discussed
in the literature, to wit energy loss, and angular deflection
 (for recent accounts see e.g. the above):
 A proton with energy above \tte\
loses energy effectively mainly by pion production on diffuse-background,
low-energy photons.
 For example, for arrival energy of $1.7\times
10^{20}$ \eV, the initial energy has to be $\sim 10^{21}$ \eV\ for an origin
distance of $D\simeq 100\,\Mpc$,
 and $\sim 10^{25}$ \eV\ for $D\simeq 200$ \Mpc.
 Such energies are very taxing for any conceivable
 acceleration mechanism. It is believed, then, that
 the distance to the sources
 of the FEE and AGASA events cannot be more than $\sim (100-200)$ \Mpc.
\par
The particles produced by the above processes, especially
 ultra-high-energy photons, may also be detected on earth in positional
or temporal correlation with GRBs. We intend to discuss these separately.
\par
The second effect is the change in the apparent direction of the
particle due to Larmour curving of the trajectory by intervening
magnetic fields. For a homogeneous field, and a constant CR energy,
 the disparity angle is
\BE \alpha\simeq {D\over 2\rl}
\simeq 0.8\left({D\over 3\,{\rm \kpc}}\right)
\left({E\over 10^{20}\,{\rm \eV}}\right)^{-1}
\left({\Bp\over 10^{-6}\,{\rm G}}\right)
\;{\rm deg}, \label{alpha}  \EE
 where $\rl=E/e\Bp$ is the gyroradius for a proton
of energy $E$, and $\Bp$ is component of the magnetic field perpendicular
to the line of sight.
In a tangled field the process is one of random walk in the angle.
Within each cell of roughly homogeneous field (of mean size $\dc$--the
reversal length) the spread in angles increases by an amount given by
eq.(\ref{alpha}) with $D=\dc$. The random-walk step increases as $E$
decreases along the trajectory, and the effect has to be integrated along
 the line of sight.
\par
In the context of the \U -GRB association
 we note that trajectory wriggling also
produces a time delay of the CR particle. Unlike the angle deviation,
which random walks, the time delay only increases with length of travel.
 For a homogeneous field, constant energy, and small departure from the
line of sight, the time delay is
\BE \Dt\simeq {1\over 24}{D^3\over c\rl\^{2}}\simeq
0.3\left({D\over 3\,{\rm \kpc}}\right)^3
\left({E\over 10^{20}\,{\rm \eV}}\right)^{-2}
\left({\Bp\over 10^{-6}\,{\rm G}}\right)^2
{}~{\rm yr}. \label{delt}  \EE
(Geometrical factors of order of a few may enter in other geometries).
In this case $\Dt$ is related to the disparity angle $\a$ by
\BE \Dt={1\over 6}{D\a^2\over c}.  \label{rela} \EE
\par
When the magnetic field is tangled,
a simple estimate of the
effect can be obtained by picturing the
line of sight to be divided into segments of mean length $\dc$
 over which the magnetic field is constant. The minimum time delay will
be produced when the particle returns to the line of sight within
each cell (the particle could wander off the line of sight and return
only upon arrival, in which case the time delay could be much larger).
Then $\Dt\_{min}\sim\int d\dt$, where
\BE d\dt\simeq
0.3\left({dD\over 3\,\kpc}\right)
\left({E\over 10^{20}\,{\rm \eV}}\right)^{-2}
\left({\Bp\over 10^{-6}\,{\rm G}}\right)^2\left({\dc\over 3\,\kpc}\right)
^2~{\rm yr} \label{ddelt}  \EE
is the delay accumulated over path length $dD$;
we assume that $\dc$ is small compared with the length
over which $E$ varies appreciably.
\par
In principle, propagation effects as discussed above may arise in
the galaxy hosting the source, in our galaxy, and in the IGM. On the
scale of a galaxy there is hardly a loss of energy, so constant
$E$ may be assumed in the above estimates. Magnetic field effects
in the host galaxy contribute negligibly to the angle disparity
(in the present context).
\par
Consider first the contribution of the Galaxy alone.
The FEE particle comes roughly from the galactic anti-centre. From
 Fig. 1a of \cite{vll91}, which shows a magnetic map of the relevant
 region, and from his Table 1, we see that it is appropriate
to take, approximately, $\Bp\sim(2-3)\mu$G and $D$ and $\dc$ of a few
kiloparsecs. With the energy of the FEE particle
eqs. (\ref{delt}) and (\ref{ddelt}) give a
time delay of a few months consistent with association the FEE with
 GRB 910503. The AGASA CR comes from galactic
 latitude of $\sim 40\deg$,
so its trajectory cuts only about 2 \kpc\ through the galactic-disk
 field--which is said to have a scale hight of $\sim 1\, \kpc$
 (\cite{k94})--but its
 energy is smaller, and we expect a similar galactic time delay
(and halo fields may also contribute). The corresponding spread in angles
is about $1\deg$.

\par
The extragalactic contribution is much more uncertain.
 In this connection
note that the CR particle is expected to alternate in identity between
a proton and a neutron due to the pion-producing interactions, so
it is only during about half of its journey that it is subject
to magnetic field effects. (The chances for identity change in the galaxy
are small, and we can ignore the effect in the above estimate.)
 Second, because the incremental time delay decreases fast with
 increasing energy, only the
end part of the intergalactic journey contributes appreciably to it
(assuming similar magnetic fields along the way). We see from Fig. 1 of
Sigl, Schramm \& Bhattacharjee (1994), which shows the decrease of
energy with travel length, that only about the last $(25-50)\Mpc$
 need be taken into account; this should be halved due to
the proton-neutron alternations.
\par
Very little is known about the intergalactic field at large.
 Kronberg (1994) cites an upper limit of $10^{-9}$ G on $B$
and of $1\,\Mpc$ on $\dc$. Assuming these values,
 and an effective propagation
distance of $25\,\Mpc$, with an energy of $3\times 10^{20}\,\eV$,
we obtain from eq.(\ref{ddelt}) a minimum
 time delay of about 25 years. Thus, confirmation
of the association, with a time delay of order of a year
or less, will cast a useful constraint on $B\dc$ for the
intergalactic magnetic field. Interestingly, in this context,
 \cite{p95} has recently
proposed to probe intergalactic magnetic fields in the range
($10^{-12}-10^{-24})$ G using the (rather shorter)
time delay of high-energy gamma-rays.
His analysis may be relevant to ultra-high-energy photons that are
produced by the \U s on their way, e.g. from $\pi^0$ decay.
\par
If the CR arrives to the galaxy as a neutron, only
propagation effects in the IGM and host galaxy contribute.
 The primary CR may, after all, be
 a neutrino, produced by a proton in the host galaxy.
As it propagates
subsequently without disturbance, it is expected to arrive
without energy loss, or angular disparity. The only contribution to
a time delay may come from a host galaxy--the neutrino being produced
by an accelerated proton, which is subject to the above magnetic-field
effects in te host galaxy. In this case, one is freed from the
energy-loss constraint on the distance. Sigl, Schramm \& Bhattacharjee
(1994) argue
that a neutrino primary can only be an atypical instance, and that
for every neutrino event there
must be many proton events with similar arrival energy.

\section{Discussion}
\par
If the GRB-\U\ association is confirmed by future correlations,
further insight into the nature of the underlying mechanism can
be gained by combining the constraints we deduce, separately,
 for the two phenomena.
 For instance, the position of a strong GRB
may be much better defined than that of an \U, and, furthermore, the
 GRB's position can be trusted as it is not affected by magnetic fields.
 In the case of
  FEE-GRB 910503 the combined $1\deg\times 1'$
 error box of the latter is some 2500 times
smaller in area than that of the former. With additional satellites,
GRB error boxes may be reduced even further by an order of magnitude
or two (as has been the case in the seventies). If it is also confirmed
that \U s are protons, the energy-loss argument greatly constrain the
distance of the source. Combined, the two constraints reduce greatly
the volume in which to look for counterparts.
\par
If the GRB event is
 also to produce \U\ s, the presence of very strong magnetic
fields will be implied.
 What little we know about
the physical conditions existing in gamma-ray
bursters during an outburst makes them highly conducive
 to the acceleration of \U.
 For example, by one class of models, GRBs are produced in
 differentially rotating disks that are formed by
a merger of a neutron-star binary (\cite{pac86};
\cite{mhim93}; \cite{pir94}), or by neutron stars that
have arisen from accretion-induced collapse (\cite{us92}). The
magnetic field, $B_{_{\rm S}}$, at the surface of these objects
may be as high as $\sim 10^{16}-10^{17}$ G (\cite{npp92};
\cite{us92}; \cite{td93}). The angular
velocity of these objects is $\Omega \sim 10^4$ s$^{-1}$. The potential
difference between the surface of such an object and infinity is
(e.g., \cite{rs75})
\BE \Delta \varphi_{\rm max}={\Omega^2B_{_{\rm S}}R^3\over 2c^2}
\simeq 1.7\times 10^{23}\left({\Omega\over 10^4\,{\rm s}^{-1}}
\right)^2\left({B_{_{\rm S}}\over 10^{16}\,{\rm G}}\right)
\left({R\over 10^6\,{\rm cm}}\right)^3\;\;{\rm V}\,, \label{dphi}\EE
\noindent where $R\simeq (1-2)\times 10^6$ cm is the radius of the
object. Charged particles that flow away from the surface
 may be accelerated, in principle, up to
the energy $E_{\rm max}\simeq e\Delta \varphi_{\rm max}$, which
is more than is needed to produce the observed \U s, at a distance of
up to $\sim100\,\Mpc$.
\par
Alternatively, particles may be accelerated by relativistic shocks
that may be formed in an unsteady relativistic wind
(\cite{rm94}). The maximum energy that
may be achieved by protons in the process
 is (\cite{ssb94} and references therein)
\BE E_{\rm max}\simeq 10^{17}\left({Br_{\rm sh}\over 0.001\,{\rm pc}
\,{\rm G}}\right)\;\;{\rm \eV}\,, \label{emax} \EE
 where $r_{\rm sh}$ is the characteristic size of the shock.
In our case
$r_{\rm sh}$ is about the distance from the compact object to the
shock. In the wind, the value of $Br_{\rm sh}$ does not depend on
the distance from the compact object, beyond the light cylinder,
 and is $\sim B_{_{\rm S}}
(\Omega R/c)^3(c/\Omega) = \Omega^2B_{_{\rm S}}R^3/c^2$.
 We can thus see that $e\Delta \varphi_{\rm max}$ with
$\Delta\varphi_{\rm max}$ from equation(\ref{dphi}) coincides with $E_
{\rm max}$ given by equation (\ref{emax}) within a factor of 2 or so.

\par
If, as has been suggested, the intrinsic luminosity of GRBs is rather
uniform, and thus fluence is strongly indicative of distance; and if,
further, the \U\ luminosity is a given fraction of the gamma luminosity,
then we expect the probability to detect a \U\ in association with
a GRB to decrease steeply with decreasing GRB fluence.
What is the total energy flux in \U s per one outburst?
Given that only one particle was observed for each of the two
GRBs we cannot, directly, estimate this quantity without knowing
the probability to detect a \U\ per GRB. We deduce that this probability,
for detectors of the size operating today, is rather smaller than
unity, even for those GRBs as energetic as GRB 910503:
If one \U\ per GRB was typical say for AGASA, with its area of about
$100\,\kmsq$ (\cite{yhh95}),
 the AGASA event would have indicated a total \U\ fluence of
$\sim 3\times 10^{-4}\,\ecmt$, at earth, compared with the total BATSE
fluence of GRB 921230 of $\sim 4\times 10^{-5}\,\ecmt$.
In addition, \U\ energy degrades in travel, so this would have given
a high ratio (perhaps as high as 100)
 of \U/GRB energy at the source. This high ratio can be avoided if
 the probability for AGASA to detect an \U\ for a GRB
as energetic as GRB 921230 is at most a few percent; this would
correspond to an energy ratio of order unity--not straining further
the energetics of the GRB source. More pertinently, this is in
keeping with the fact that a few tens of GRBs as strong as
GRB 921230 are observed per year,
 with only one \U\ by AGASA over a few years.
 GRB 910503 is about seven times more energetic, and about a few tens
 times rarer, so we expect on such burst in a few to be detected
in \U.
 In both cases, detection statistics is consistent
 with roughly similar energies going into \U s, and gamma-rays.

\par
 A remarkable gamma-ray burst, GRB 940217, was observed on
 Feb 17 1994. The total fluence above 20 keV of $\simeq
6\times 10^{-4}\,\ecmt$ is twice higher than that of
 GRB 910503. A photon with the energy of 18 GeV
was detected one-and-a-half hour after the BATSE burst trigger from
the region of GRB 940217 (\cite{hdm94}). This is the highest
energy of photons that was ever observed in a burst.
 The direction is known quite well:
a combined COMPTEL--EGRET--Ulysses error box of only $\sim
0.5\deg \times 0.1\deg$. This burst is very promising as a source
of \U s.

\end{document}